# Atomic-scale Frustrated Josephson Coupling and Multi-condensate Visualization in FeSe

Nileema Sharma[1,2]†, Matthew Toole[1,2]†, James McKenzie[1,2], Sheng Ran[3], and Xiaolong Liu[1,2]*

[1]Department of Physics and Astronomy, University of Notre Dame, Notre Dame, IN 46556, USA

[2]Stavropoulos Center for Complex Quantum Matter, University of Notre Dame, Notre Dame, IN 46556, USA

[3]Department of Physics, Washington University in St. Louis, St. Louis, MO 63130, USA

† These authors contributed equally to this work.

* Corresponding author. Email: xliu33@nd.edu

**In a Josephson junction involving multi-band superconductors, competition between inter-band and inter-junction Josephson coupling gives rise to frustration and spatial disjunction of superfluid densities among superconducting condensates[1,2,3,4,5,6,7]. Such frustrated coupling manifests as quantum interference of Josephson currents from different tunneling channels and becomes tunable if channel transparency can be varied[5,6,7,8]. To explore these unconventional effects in the prototypical $s^{\pm}$-wave superconductor FeSe[9], we use atomic-resolution scanned Josephson tunneling microscopy (SJTM)[10,11,12,13] for condensate-resolved imaging and junction tuning—capabilities unattainable in macroscopic Josephson devices with fixed characteristics. We quantitatively demonstrate frustrated Josephson tunneling by examining two tunneling inequalities. The relative transparency of two parallel tunneling pathways is found tunable, revealing a tendency towards a 0-$\pi$ transition with decreasing SJTM junction resistance. Simultaneous visualization of both superconducting condensates reveals anti-correlated superfluid modulations, highlighting the role of inter-band scattering. Our study establishes SJTM as a powerful tool enabling new research frontiers of multi-condensate superconductivity.**

Multi-band superconductivity develops in materials with overlapping energy bands at the Fermi level. Each condensate is characterized by a macroscopic wavefunction $\psi_i = |\psi_i|e^{i\varphi_i} \equiv \sqrt{n_i}e^{i\varphi_i}$ with phase $\varphi_i$ and mean-field superfluid density $n_i$. Simultaneous intra- and inter-band pairing interactions can lead to rich phenomenology and exotic states such as Leggett's collective mode[4], chiral superconductivity[14,15,16], type-1.5 superconductivity[17,18], phase solitons[19], and vortex fractionalization[20,21].

In a Josephson junction formed between an *s*-wave and an *N*-band superconductor, frustration takes place if inter-band and inter-junction Josephson coupling competes (i.e., $\varphi_i$ are not identical on all bands), leading to quantum interference of Josephson current of parallel tunneling channels in the band basis[1,2,3,4,5,6,7,8], of which the direct detection has been elusive. Compared to conventional *s*-wave junctions, the total Josephson critical current $I_J$ becomes a phase-sensitive sum:
$$I_J = \sum_{i=1}^N I_{J,i} \cos\chi_i \tag{1}$$
where $I_{J,i} > 0$ and $\chi_i = \varphi_i - \varphi_S$ are the Josephson critical current and gauge-invariant phase difference between band *i* and the *s*-wave electrode under zero magnetic field[1,3,5,6]. For an SJTM junction, this interference effect is schematically shown in Fig. 1a where a superconductive *s*-wave tip has a gap magnitude of $\Delta_T$. The critical Josephson current $I_{J,i}$ and junction resistance $R_{N,i}$ of each tunneling channel can be described by the asymmetric Ambegaokar-Baratoff (A-B) formula[6]
$$I_{J,i} R_{N,i} = \frac{2\Delta_T |\bar{\Delta}_i|}{e(\Delta_T + |\bar{\Delta}_i|)} K\left(\left|\frac{\Delta_T - |\bar{\Delta}_i|}{\Delta_T + |\bar{\Delta}_i|}\right|\right) \equiv \frac{\pi \Delta_{\text{eff},i}}{2e} \tag{2}$$
where $\bar{\Delta}_i \equiv \langle \Delta_i(\boldsymbol{k}) \rangle > 0$ is the superconducting gap magnitude of band *i* of the multi-band superconductor averaged over the Fermi surfaces, and $K(x)$ the complete elliptic integral. $\Delta_{\text{eff},i}$ can be taken as the gap magnitudes of effective symmetric junctions. Following Ref. 3, if we define $\Delta_{\text{eff,min}} = \min(\Delta_{\text{eff},i})$ and $\chi_i = 0$ for $i = 1\sim N$ (i.e., *s*-wave), the lower threshold of $I_J R_N$ for a multi-band *s*-wave Josephson junction can be obtained
$$I_J R_N = \frac{\pi R_N}{2e} \sum_{i=1}^N \frac{\Delta_{\text{eff},i}}{R_{N,i}} \geq \frac{\pi}{2e} \Delta_{\text{eff,min}} \tag{3}$$
where we have used the total junction resistance[3,4]
$$R_N = 1/\sum_{i=1}^N \frac{1}{R_{N,i}} \tag{4}$$
Therefore, the effect of frustration and Josephson interference must be manifested as
$$I_J R_N < \frac{\pi}{2e} \Delta_{\text{eff,min}} \tag{5}$$

In the simplest case, frustration happens between an s-wave superconductor interacting with a sign-changing $s^\pm$-wave superconductor with negative inter-band Josephson coupling. Furthermore, depending on the relative inter-junction coupling strengths between the *s*-wave superconductor and the two bands, the junction may display a negative Josephson critical current (i.e., a π-junction) and a 0-π transition if the junction transparencies of each channel can be tuned[5,6,7,8,22]. In the limit of strong frustration when the Josephson coupling energies in the two channels are fully compensated, the relative phases of the order parameters across the junction can deviate from 0 or π, giving rise to time-reversal symmetry breaking and $\varphi$-junctions[1,2,6,8]. While it is generally accepted that $s^\pm$-wave pairing can be found in materials including Fe-based superconductors[9], to our knowledge, direct observation of Josephson tunneling interference due to frustration has not been demonstrated, possibility due to compromised interfaces during junction fabrication and spatial heterogeneities. Most critically, testing of the Eqn. 5 requires faithful extraction of $I_J$ and all superconducting gaps with highest resolution, which is technically challenging when multi-gaps are similar in energy. In addition, the existence of more than one superconducting condensate suggests an internal degree of freedom in the inter-condensate spatial arrangement of superfluid densities, which can reveal the relative significance of inter-band scattering and Josephson coupling. Yet, visualizing multiple superconducting condensates at the atomic scale remains unachieved because such a technique does not exist.

Here, we leverage atomic resolution SJTM at 0.3 K to provide quantitative evidence of Josephson tunneling interference between Nb and FeSe as a result of frustrated Josephson coupling and reveal a spatially anti-correlated superfluid density modulations via atomic scale condensate-resolved visualization. Such capability is enabled by the drastically improved energy resolution and simultaneous Josephson and quasiparticle visualization using SJTM[10,13,23,24]. FeSe is a prototypical $s^\pm$-wave superconductor with electronic nematicity and orbital-selective Cooper pairing of electrons preferentially from $d_{yz}$ orbitals of Fe[9,25], manifesting twofold symmetric superconducting gaps $\Delta_{1,2}(\mathbf{k})$ on the hole (at Γ) and electron (at X) pockets with opposite signs, respectively (Figs. 1b,c). A representative topographic image of FeSe taken with a Nb tip is shown in Fig. 1d, where a twin boundary separates two nematic domains featuring dumbbell-shaped defects orientated in orthogonal directions. Due to the gap of the s-wave Nb tip ($\Delta_T \approx 1.2$ meV), differential conductance ($dI/dV$) spectra taken at varying tunneling junction resistance ($R_N$, Fig. 1e) clearly reveal two quasiparticle coherence peaks at $\Delta_{1,m} + \Delta_T$ and $\Delta_{2,m} + \Delta_T$, where $\Delta_{1,m} \approx 2.3$ meV and $\Delta_{2,m} \approx 1.5$ meV are the respective maximum gaps on the hole and electron pockets of FeSe (ref. 25). Simultaneously visualized Cooper-pair tunneling current is shown in Fig. 1f. In the Ivanchenko-Zil'Berman (I-Z) theory[26] of phase-diffusive Josephson junctions that applies to voltage-biased SJTM[10,11,12,13,23,24], the pair-current is

$$I(V) = \frac{1}{2} I_J^2 Z V/(V^2 + V_c^2) \qquad (6)$$

such that the maximum Josephson current $I_m = \frac{Z}{4V_c} I_J^2$ is observed at $V = V_c$. Here, $Z$ is the total electromagnetic impedance of all elements and circuitry adjacent to the junction. Together with the standard A-B relation for s-wave junctions, $I_J R_N = \frac{\pi\Delta}{2e} \tanh\left(\frac{\Delta}{2k_B T}\right)$, we anticipate a linear relationship of

$$I_J \propto \sqrt{I_m} \propto 1/R_N \tag{7}$$

which is indeed observed in *s*-wave SJTM junctions of NbSe$_2$-Nb (refs. 13, 23), Pb-Pb (refs. 27, 28), and Nb-Nb (Supplementary Fig. 1). However, in the case of FeSe-Nb, deviation from this linear relationship is clearly and reproducibly observed in Fig. 1g and Supplementary Fig. 2. To understand this, we can take $\chi_1 = 0$ and $\chi_2 = \pi$ considering the $s^\pm$-wave pairing of FeSe, and obtain the $I_J R_N$ product from Eqn. 1 as

$$I_J R_N = \frac{\pi}{2e} \frac{\Delta_{\text{eff},1} - \epsilon \Delta_{\text{eff},2}}{\epsilon + 1} \equiv \frac{\pi}{2e} \Delta_{\text{eff}} \tag{8}$$

Here, we have defined $\epsilon \equiv R_{N,1}/R_{N,2}$ and $\Delta_{\text{eff}} \equiv \frac{2e}{\pi} I_J R_N$. As we will demonstrate later (Fig. 3h), $\epsilon$ increases as $1/R_N$ increases. This leads to decreasing $\frac{\Delta_{\text{eff},1} - \epsilon \Delta_{\text{eff},2}}{\epsilon + 1}$ and therefore the sublinear increase of $\sqrt{I_m} \propto I_J$ as predicted by Eqn. 8 and observed in Fig. 1g. Furthermore, the slower increase of $I_J$ suggests a decreasing $I_J R_N$ product. This is consistent with approaching a 0-$\pi$ junction transition (Fig. 3i), which requires interfering tunneling pathways with opposite current flow directions. Thus, such sublinear scaling constitutes the first experimental signature of Josephson tunneling interference due to frustrated Josephson coupling.

Because $\Delta_{\text{eff},2} < \Delta_{\text{eff},1}$ in FeSe-Nb junctions, a quantitative piece of evidence of Josephson tunneling interference in $s - s^\pm$ junctions would be $\frac{\Delta_{\text{eff}}}{\Delta_{\text{eff},2}} < 1$ according to Eqns. 5 and 8. Furthermore, because $I_J R_N = \frac{\pi \Delta_{\text{eff}}}{2e} = \frac{\pi R_N}{2e} \left( \frac{\Delta_{\text{eff},1}}{R_{N,1}} - \frac{\Delta_{\text{eff},2}}{R_{N,2}} \right)$, one expects $\frac{\Delta_{\text{eff}}}{\Delta_{\text{eff},1}} < 1$. To test these hypotheses, we performed both Cooper-pair and quasiparticle tunneling spectroscopy to visualize locally the coherence peaks and Josephson tunneling current concurrently. Figure 2a is a topographic image of a defect-free area with simultaneously imaged $I_m(\boldsymbol{r})$ shown in Fig. 2b, allowing the extraction of $I_J(\boldsymbol{r})$ using internal calibrations with a Nb-Nb junction (Supplementary Note 1 and Supplementary Fig. 1). To extract $\Delta_{\text{eff},i}$, we deconvolute energy-symmetrized $dI/dV$ spectra measured in the FeSe-Nb junction to obtain the density of states (DOS) of FeSe (Supplementary Note 2). Then $\Delta_{\text{eff},i}$ is obtained via fitting the deconvoluted FeSe DOS (see an exemplary fitting in Fig. 2c) with realistic twofold symmetric gaps for both the electron and hole pockets (Supplementary Note 3 and Supplementary Figs. 3-6). Critically, compared to measuring the DOS of FeSe with conventional metallic tips, the superconductive SJTM tip gaps out the thermal smearing at the Fermi level, enabling an energy resolution beyond the Fermi-Dirac limit in such superconductor-insulator-superconductor junctions (Supplementary Fig. 7)[10,13,23,24,29]. By fitting the slope of $I - V$ curves at energies far outside the gaps[13,23,24], the junction resistance $R_N(\boldsymbol{r})$ (Fig. 3f) and $\Delta_{\text{eff}}$ are determined (discussions on the influence of $R_N$ are given in Supplementary Note 4 and Supplementary Fig. 8). With high fitting fidelity (Supplementary Figs. 5, 9), Figures 2d-f show the extracted $\Delta_{\text{eff}}(\boldsymbol{r})$, $\Delta_{\text{eff},1}(\boldsymbol{r})$, and $\Delta_{\text{eff},2}(\boldsymbol{r})$, respectively, in the same field of view (FOV) as Fig. 2a, displaying long range undulations possibly due to sub-surface heterogeneities. Histograms of those effective gaps are shown in Supplementary Fig. 10. The effective gap ratios $\lambda_2 \equiv \frac{\Delta_{\text{eff}}}{\Delta_{\text{eff},2}}$ and $\lambda_1 \equiv \frac{\Delta_{\text{eff}}}{\Delta_{\text{eff},1}}$ are visualized in Figs. 2g,h, respectively. From the histograms (Fig. 2i), clearly both indices are smaller than unity, thus providing strong quantitative evidence of atomic scale Josephson tunneling interference between FeSe and Nb due to the sign-changing gaps of FeSe.

In SJTM, the directly measurable quantities are the overall Josephson current and junction resistance, while generally, the channel-specific $I_{J,i}$ and $R_{N,i}$ for an $N$-band superconductor cannot be retrieved. However, the case of $N = 2$ (e.g., FeSe) is exactly solvable: combing Eqns. 1, 2 and 4, the four unknowns can be determined as

$$I_{J,1} = \frac{\pi}{2e} \frac{\Delta_{\text{eff},1}(\Delta_{\text{eff},2}+I_J R_N)}{R_N(\Delta_{\text{eff},1}+\Delta_{\text{eff},2})}, \qquad I_{J,2} = \frac{\pi}{2e} \frac{\Delta_{\text{eff},2}(\Delta_{\text{eff},1}-I_J R_N)}{R_N(\Delta_{\text{eff},1}+\Delta_{\text{eff},2})} \tag{9}$$

$$R_{N,1} = \frac{(\Delta_{\text{eff},1}+\Delta_{\text{eff},2})}{\Delta_{\text{eff},2}+I_J R_N} R_N, \qquad R_{N,2} = \frac{R_N(\Delta_{\text{eff},1}+\Delta_{\text{eff},2})}{\Delta_{\text{eff},1}-I_J R_N} R_N \tag{10}$$

Such channel-resolved visualization is made possible by simultaneous Cooper-pair and quasiparticle tunneling to measure $I_J$, $\Delta_{\text{eff},i}$, and $R_N$ at the same atomic location. Figures 3a-f show the extracted $I_{J,i}(\boldsymbol{r})$ and $R_{N,i}(\boldsymbol{r})$, as well as measured $I_J(\boldsymbol{r})$ and $R_N(\boldsymbol{r})$ in the same FOV as in Fig. 2. We note that solving Eqns. 1, 2 and 4 by assuming $s^{++}$-wave pairing ($\chi_1 = \chi_2 = 0$) leads to unphysical negative junction resistance and critical current magnitude (Supplementary Figure 11), thus supporting the robustness of our model. Using spatial coordinates as an implicit variable in Fig. 3g, $\varepsilon = R_{N,1}/R_{N,2}$ is found to be anti-correlated with $R_N$ (Fig. 3h), explaining the sub-linear relationship observed in Fig. 1g based on Eqn. 8. This suggests a $R_N$-tunable relative coupling between the Nb tip to the two bands of FeSe, which cannot be realized in macroscopic Josephson devices with fixed junction geometries. As shown in Fig. 3i, the normalized Josephson critical current ($I_J R_N$) extracted in the same FOV decreases as a function of $\varepsilon$, suggesting a tendency towards a $0$-$\pi$ transition[5,6,7,8,22] as predicted from Eq. 8. However, reaching such a transition is currently limited by the increased SJTM junction instability at low $R_N$, which might be alleviated by the use of blunter tips sacrificing spatial resolution and non-layered $s^{\pm}$-wave superconductors.

For single-band (or effectively single-band) $s$-wave superconductors, recent SJTM works[12,13,23,24] have demonstrated unique capabilities of visualizing superfluid densities $n(\boldsymbol{r}) \propto (I_J R_N)^2$ at the atomic scale via Josephson tunneling, revealing spatially modulated superconductivity from chemical heterogeneities, Cooper-pair density waves (PDWs) or quantum vortices. Here, we explore the possibility of condensate-resolved visualization of superfluid densities in FeSe. For a multi-band superconductor, one has[12]

$$I_J R_N \propto \sum_{i=1}^{N} \sqrt{n_i} \cos\chi_i \tag{11}$$

Therefore, based on Eqn. 1, the superfluid density of each condensate can be measured as

$$n_i(\mathbf{r}) \propto \left(I_{J,i} R_N\right)^2 \propto \left(\frac{R_N \Delta_{\text{eff},i}}{R_{N,i}}\right)^2 \quad (12)$$

where each quantity can be obtained from channel-resolved Josephson tunneling measurements (Eqns. 9. 10, and Fig. 3). In Figs. 4a,b, we show $n_i(\mathbf{r})$ measured in the same FOV as in Fig. 2 and the corresponding Fourier transforms $n_i(\mathbf{q})$ with the Bragg peaks and $\mathbf{q} = 0$ circled in blue and red, respectively. Spatial undulations of $n_i(\mathbf{r})$ are clearly seen and better visualized in the inverse Fourier transforms $n_i'(\mathbf{q})$ in Figs. 4c,d using the $\mathbf{q} \sim 0$ components. Both $n_1'(\mathbf{q})$ and $n_2'(\mathbf{q})$ exhibit domains of very different magnitudes (dashed curves are a guide to the eye), yet appear spatially anti-correlated. Indeed, such anti-correlation is confirmed by a cross-correlation coefficient of –0.72 between Figs. 4a and 4b (Supplementary Fig. 12) and statistically confirmed in Fig. 4e.

To provide theoretical insight into this observation, we employ the phenomenological Ginzburg-Landau (G-L) theory. For simplicity, we assume superconducting order parameters $\psi_i$ ($i = 1, 2$) of two condensates in the form of $\psi_i = \sqrt{n_i} e^{i\varphi_i} = [A_i + \delta A_i \cos(qx + \theta_i)] e^{i\varphi_i}$, where a small but spatially modulated 1D component with wavevector $q$ is added on top of a large uniform background ($\delta A_i \ll A_i$). For convenience, we can define $\theta_1 = 0$ and $\theta_2 = \theta$. Because the superfluid densities $n_i(x) = |\psi_i(x)|^2$, the leading order spatial phase difference between $n_1$ and $n_2$ is $\theta$ as schematically shown in the inset of Fig. 4f. The general two-component G-L free energy density functional is[30,31]

$$F = F_1 + F_2 + F_{12} \quad (13)$$

Here, $F_i$ ($i = 1,2$) represents the free energy density of each condensate up to fourth order

$$F_i = \alpha_i |\psi_i|^2 + \frac{1}{2}\beta_i |\psi_i|^4 + \frac{1}{2}(\nabla \psi_i)(\nabla \psi_i)^* \quad (14)$$

$F_{12}$ represents inter-condensate coupling involving inter-band Josephson coupling ($\eta_1$ term), gradient coupling from inter-band scattering ($\nu$ term), and a fourth order term ($\eta_2$ term)

$$F_{12} = -\frac{1}{2}\eta_1(\psi_1^* \psi_2 + \psi_1 \psi_2^*) - \frac{\nu}{2}[(\nabla \psi_1)^*(\nabla \psi_2) + (\nabla \psi_1)(\nabla \psi_2)^*] + \eta_2 |\psi_1|^2 |\psi_2|^2 \quad (15)$$

For $s^\pm$-wave pairing, the inter-band Josephson coupling has $\eta_1 < 0$ and the phase difference of the order parameters is $\varphi_2 - \varphi_1 = \pi$. Then, the free-energy per unit length can be obtained as

$$\bar{F} \equiv \frac{q}{2\pi} \int_0^{2\pi/q} F dx = \frac{1}{2} \delta A_1 \delta A_2 (\nu q^2 + \eta_1) \cos(\theta) + C \quad (16)$$

where we have ignored a higher order term $\sim \eta_2 (\delta A_1 \delta A_2)^2 \cos(2\theta)$ and $C$ is a constant. This suggests the inter-condensate spatial correlation is controlled by the interplay between Josephson coupling ($\eta_1$) and inter-band scattering ($\nu$). With $\nu = 0$, $\bar{F}$ will be minimized at $\theta = 0$. Therefore, the observed nearly anti-correlated superfluid modulation ($\theta \approx \pi$) in FeSe suggests not only non-zero but also sizable inter-band scattering such that $\nu q^2 + \eta_1 > 0$. Because $\nu q^2$ scales with $q^2$, such anti-correlations will be reinforced at small length scales in real space, which is indeed observed (Supplementary Fig. 12).

In summary, using atomic resolution SJTM for simultaneous visualization of Josephson tunneling and superconducting gaps with superior energy resolution, we have detected quantitative evidence of frustrated Josephson coupling between a Nb tip and FeSe in agreement with theory expectations. We have further demonstrated condensate-resolved superfluid imaging, revealing a spatially anti-correlated modulation between the two condensates that can be rationalized by the interplay of inter-band Josephson coupling and scattering using the G-L theory. The unique experimental capabilities will enable exciting new research frontiers of exotic multi-band superconductivity including visualizing condensate-resolved PDWs and inter-condensate PDW phase/amplitude correlations, as well as possibly disparate behaviors of individual condensate near vortices, domain walls, and boundaries.

**Acknowledgements**
The authors thank D. Morr, B. Janko, M. Eskildsen, and Y.-T. Hsu for valuable discussions. This work is supported by the U.S. Department of Energy, Office of Science, Office of Basic Energy Sciences (DE-SC0025021 to X.L. for advanced SJTM measurements of unconventional superconductivity), the National Science Foundation, Division of Materials Research (DMR-2236528 to S.R.), startup funding from the University of Notre Dame and the Stavropoulos Center for Complex Quantum Matter (to X.L. for initially setting up the research laboratory), and the Notre Dame Materials Science and Engineering Fellowship (to N.S. and M.T.).

**Author contributions**
X.L. conceived the project. N.S. and M. T. performed the measurements. S. R. provided the FeSe crystals. N.S., M. T., and J.M. performed data analysis. X.L. wrote the manuscript with input from all authors. All authors contributed to data interpretation.

**Competing Interests**
The authors declare no competing interests.

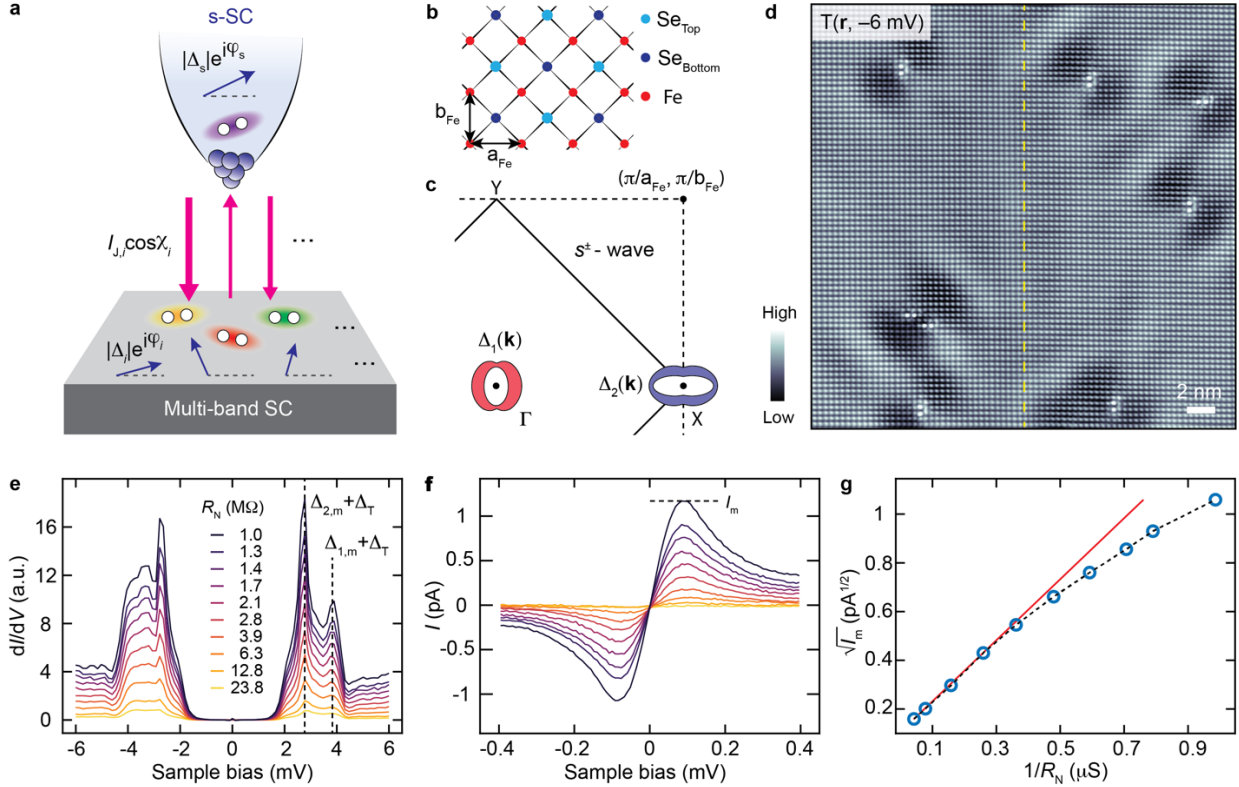

**Figure 1. Frustrated Josephson tunneling in multi-band superconductors and FeSe.**
**a**, Schematic of Josephson tunneling interference between an *s*-wave superconductor (s-SC) and a multi-band superconductor due to the competition between inter-condensate and inter-junction phase differences. **b**, Schematic of FeSe crystal structure viewed from the out-of-plane direction. Orthorhombic distortion at low temperature leads to slight differences between the lattice constants ($a_{Fe} \neq b_{Fe}$). **c**, Schematic of the Fermi surface and Brillouin zone of FeSe. The solid and dashed lines indicate the two-Fe and one– Fe Brillouin zones, respectively. A hole and electron pocket both with twofold symmetric superconducting gaps $\Delta_{1,2}(\mathbf{k})$ but opposite signs (i.e., shaded red and purple) are located respectively at the Γ and X points. **d**, Topographic image of FeSe at 0.3 K showing a nematic twin boundary (setpoint: $V_s$ = −6 mV, $I_0$ = 5.5 nA). **e**, Quasiparticle tunneling spectra between a Nb tip and FeSe at different normal state resistance $R_N$. The two coherence peaks are indicated. **f**, Josephson tunneling spectra between a Nb tip and FeSe at different $R_N$. **g**, The dependence of $\sqrt{I_m}$ on $1/R_N$ strongly deviates from linearity.

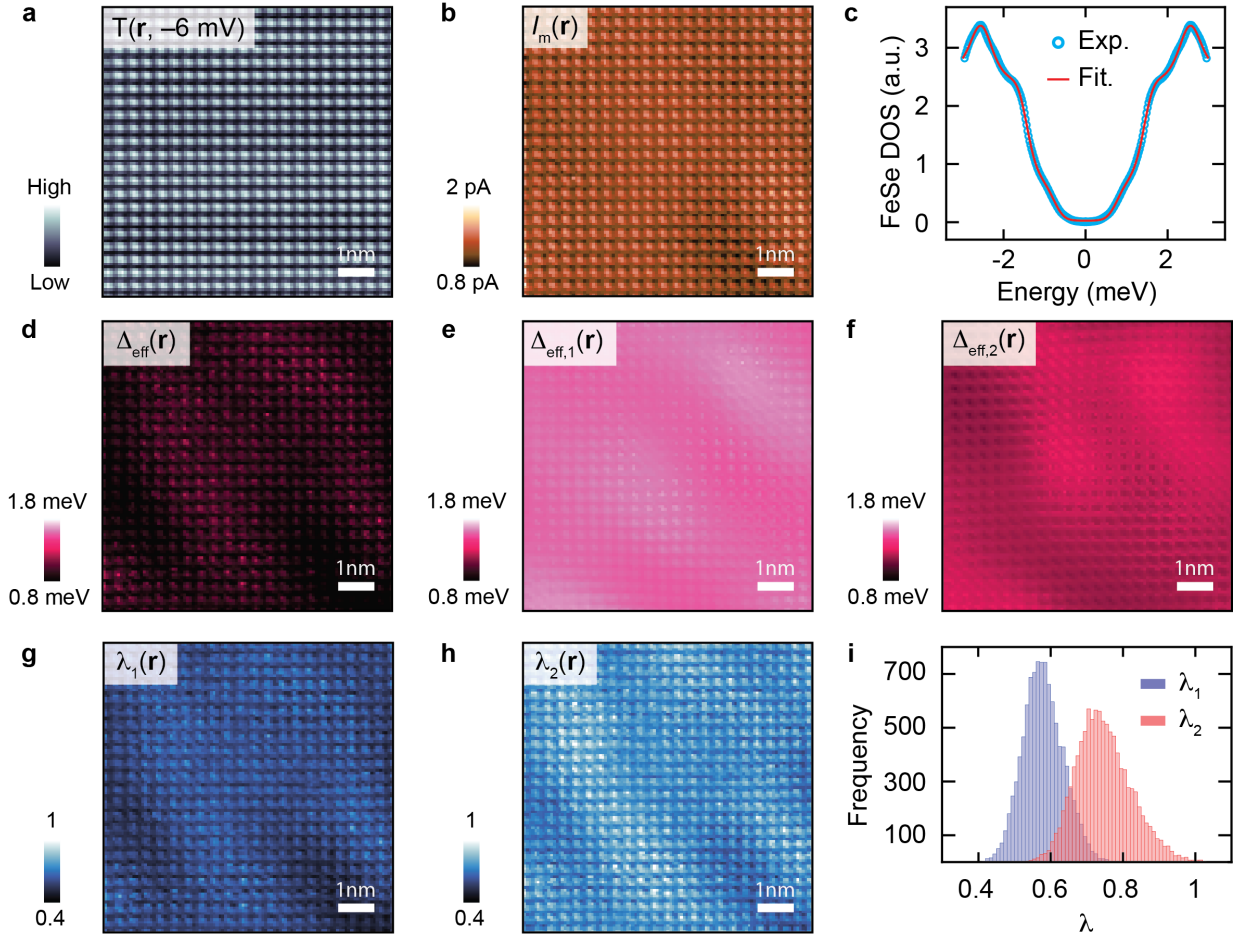

**Figure 2. Visualizing Josephson tunneling interference between Nb and FeSe.**
**a**, Topographic image of FeSe (setpoint: $V_s$ = −6 mV, $I_0$ = 9 nA). **b**, Maximum Josephson current $I_m$ visualized in the same FOV as in (a). **c**, Representative FeSe DOS deconvoluted from symmetrized experimental Nb-FeSe tunneling spectrum (blue circles) and a fitting curve using realistic anisotropic FeSe gap structures (red, with $\Delta_1^m$ = 2.55 meV, $\Delta_2^m$ = 1.56 meV, $a_1 = 0.213$, $a_2 = 0.263$; the definitions of those parameters are given in Supplementary Note 3). Over the entire map, the fitting parameters vary slightly and statistically we have $a_1 = 0.215 \pm 0.013$, $a_2 = 0.264 \pm 0.023$. **d**, Effective gaps $\Delta_{eff}(\mathbf{r})$, **e**, $\Delta_{eff,1}(\mathbf{r})$, and **f**, $\Delta_{eff,2}(\mathbf{r})$ in the same FOV as in (a). **g**, Extracted indices of $\lambda_1(\mathbf{r})$ and **h**, $\lambda_2(\mathbf{r})$ in the same FOV as in (a). **i**, Histograms of $\lambda_1(\mathbf{r})$ and $\lambda_2(\mathbf{r})$ in (g,h). Both indices being smaller than one strongly suggests interference of Josephson tunneling current between Nb and FeSe.

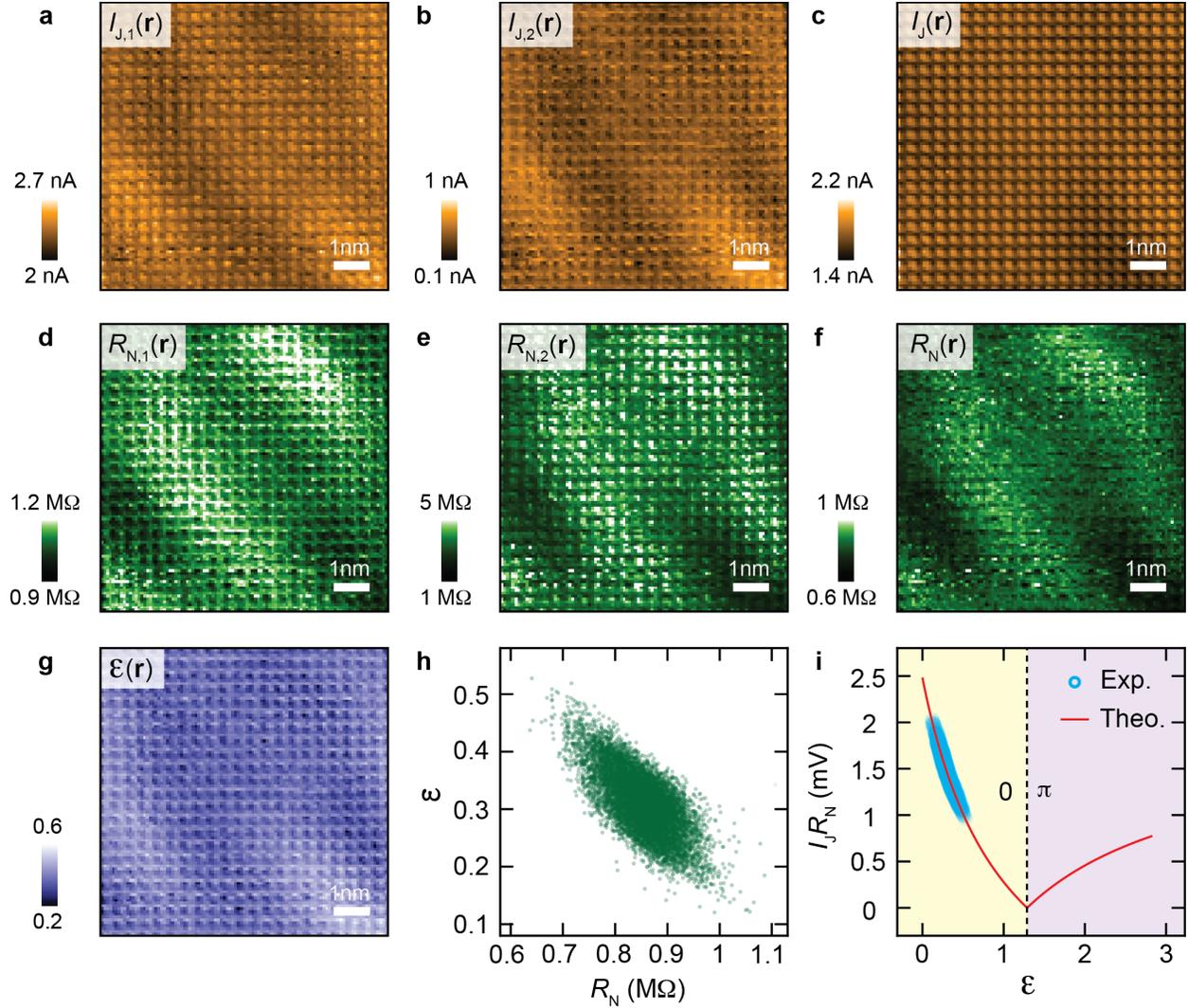

**Figure 3. Tunable and condensate-resolved Josephson tunneling.**
**a**, Channel-resolved Josephson critical current $I_{J,1}$, **b**, $I_{J,2}$, and **c**, total Josephson critical current $I_J$ visualized in the same FOV as in Fig. 2a. **d**, Channel-resolved junction resistance $R_{N,1}$, **b**, $R_{N,2}$, and **c**, overall junction resistance $R_N$ in the same FOV. **g**, Spatial variation of the ratio $\varepsilon(\mathbf{r}) = R_{N,1}/R_{N,2}$. **h**, The ratio $\varepsilon$ is anti-correlated with the junction resistance $R_N$. **i**, Plotting the normalized Josephson critical current $I_J R_N$ measured in the same FOV as a function of $\varepsilon$ and the theoretical relation (red curve) showing a 0-$\pi$ transition.

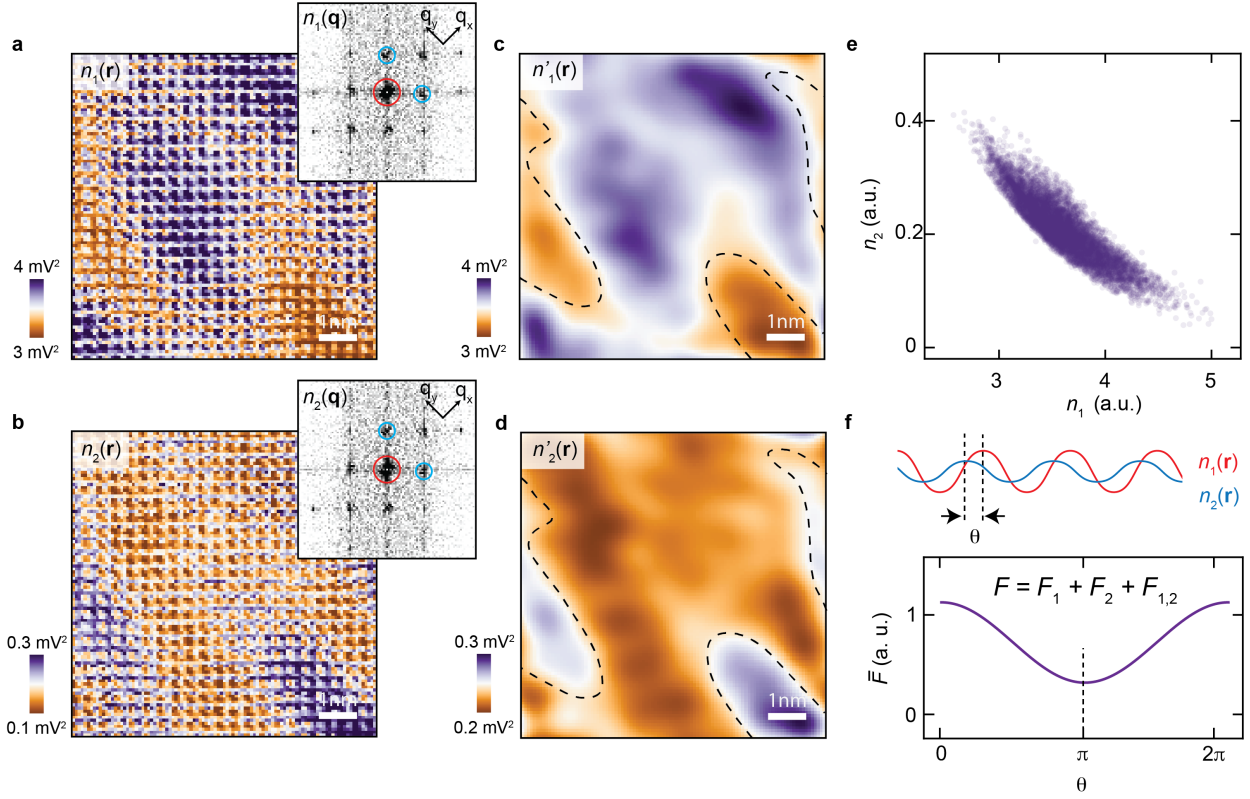

**Figure 4. Anti-correlated inter-condensate superfluid density modulation.**
**a**, **b**, Superfluid density of $n_1(\mathbf{r})$ and $n_2(\mathbf{r})$ extracted from the same FOV as in Fig. 2a. The insets are Fourier transforms where the Bragg peaks and $\mathbf{q}=0$ are indicated by blue and red circles, respectively. The cross-correlation between them is −0.72. **c**, **d**, Inverse Fourier transform of the $n_1(\mathbf{r})$ and $n_2(\mathbf{r})$ using the low-$\mathbf{q}$ components within the red circles in the insets of (a,b). An anti-correlated spatial modulation between $n'_1(\mathbf{r})$ and $n'_2(\mathbf{r})$ is clearly seen. The dashed lines are a guide to the eye. **e**, Anti-correlation between the two superfluid densities visualized by plotting $n_1(\mathbf{r})$ and $n_2(\mathbf{r})$ using $\mathbf{r}$ as an implicit variable. **f**, Schematic of free energy per unit length (Eqn. 16) plotted as a function of inter-condensate spatial phase difference $\theta$ (schematically defined in the inset) showing a minimum free energy at $\theta = \pi$.

**Methods**

SJTM experiments are performed on a Unisoku USM1300J systems at a base temperature of 0.3 K. Single crystals of FeSe are synthesized via a hydrothermal method detailed in a previous study[32] and mechanically cleaved at a temperature of ~ 80 K under ultrahigh vacuum (1 × 10$^{-10}$ Torr) before loaded into the STM head. Mechanically cut Nb tips from pure Nb wires are heated in the vacuum before being used as SJTM tips. Calibration of the superconducting gaps of Nb tips and Josephson critical currents is performed on a Nb(100) single crystal (Princeton Scientific) prepared by repeated Ar$^+$ ion sputtering (2 keV) and annealing up to 800°C. FEMTO preamplifiers are used for tunneling current amplification. Data acquisition is performed using SPECS Nanonis electronics. Spectroscopic measurements of $\frac{dI}{dV} - V$ and $I - V$ are performed using a built-in lock-in amplifier in Nanonis with a typical modulation of 30 μV and frequency of 983.7 Hz. The modulation is typically turned off for bias voltages in the range of –300 to 300 μV for the acquisition of Josephson tunneling current without external energy smearing. Tunneling spectra with varying tunneling resistance (e.g., Figs. 1e,f) are acquired by varying the tunneling current setpoint while the sample bias voltage setpoint remains fixed, effectively adjusting the tip-sample distance. MATLAB and Gwyddion are used for data processing. Details of theoretical analysis and modeling are provided in Supplementary Notes 1-5.

**Methods-only References**